\documentclass{elsart}

\usepackage{graphicx}
\usepackage{epsf}
\usepackage{wrapfig}
\usepackage{here}
\usepackage{amssymb}

\begin{document}

\runauthor{}

\begin{frontmatter}

\title{{\sc STEREO ARRAY} of 30~m Imaging Atmospheric \v{C}erenkov
Telescopes: A Next-Generation Detector for Ground-Based High
Energy Gamma-ray Astronomy}

\author{A. Konopelko}
\address{Max-Planck-Institut f\"ur Kernphysik, D-69029 Heidelberg,
Germany\\
Institut f\"ur Physik, Humboldt-Universit\"at zu Berlin,\\
Newtonstr. 15, D-12489 Berlin, Germany}

\begin{abstract}
The construction of the H.E.S.S. ({\it High Energy Stereoscopic
System}), a superior system of four 12~m imaging atmospheric
\v{C}erenkov telescopes, has been completed recently in the Namib
desert close to Windhoek, at 1800~m above sea level. This
 new generation of ground-based gamma-ray detectors has an energy
threshold of about 100~GeV in observations at zenith and a
sensitivity of about 1\% Crab flux for a point-like $\gamma$-ray
source. Such high sensitivity was achieved due to $0.1^\circ$
angular resolution, as well as, severe cosmic ray background
rejection acquired from multi-fold imaging of individual
atmospheric showers. In addition H.E.S.S. has a rather good energy
resolution of 15\%. H.E.S.S. has been taking routine observations
of the $\gamma$-ray sources since December~2003.

Similar stereoscopic arrays are currently under construction at
Kitt Peak, Arizona, and at Woomera, Australia. Two telescopes of
somewhat larger size of 17~m diameter are being built by the MAGIC
collaboration on the Canary Island of La Palma. The first of these
two telescopes has been taking data since the Fall of 2003.

The outstanding physics results achieved with H.E.S.S. already in
a first year of its exploitation strongly encourage further
development of imaging atmospheric \v{C}erenkov technique for
high-quality $\gamma$-ray observations, which is basically driven
by father reduction of the energy threshold of a forthcoming major
future $\gamma$-ray detector. Here we are dealing with such a
detector, which may allow us to achieve an energy threshold as low
as 10~GeV given a unique sensitivity of about $2 \times
10^{-13}\,\,\rm erg\, cm^{-2}\, s^{-1}$.

Basic results on performance and sensitivity for a single
stand-alone 30 m imaging atmospheric \v{C}erenkov telescope, as
well as for a system of two and five \v{C}erenkov telescopes,
derived from appropriate Monte Carlo simulations, are discussed
here.

\end{abstract}

\end{frontmatter}

\section{Introduction}

Development of further instrumentation in the field of very high
energy (VHE) $\gamma$-ray astronomy is primarily motivated by the
physics goals as perceived by the astrophysics community today
(Weekes 2004). Among those one has to mention (i) observations of
the supernova remnants (SNR), which are the conjectural sources of
VHE $\gamma$ rays; (ii) continuous studies of the physics of the
relativistic jets in active galactic nuclei (AGN); (iii) further
investigations of morphology and spectra of $\gamma$ rays from
pulsar wind nebulae (PWN); (iv) the widening of the search for
sources of pulsed $\gamma$-ray emission in VHE $\gamma$-ray band,
etc. Such variety of physics enquiry can not be contented with
only a single-type ground-based $\gamma$-ray instrument. Foremost
the physics diversity of $\gamma$-ray emission mechanisms requires
the essential observations appropriated in slightly different
energy ranges. Thus for instance further observations of AGN and
Pulsars necessitate the reduction of an instrumental energy
threshold down to, at least, 10~GeV, whereas for detection of SNR
assemblage a noticeable upgrade of sensitivity above 100~GeV is
favored. Ultimately, the design of a major ground-based
\v{C}erenkov facility for future dedicated $\gamma$-ray
observations has to conform to many requirements defined by
peculiar energy spectral shapes, various angular extents, and
strongly variable photon rates for the sources of an entirely
different nature.

\section{Basic Parameters of the Telescope Design}

The sensitivity of imaging atmospheric \v{C}erenkov telescope
(IACT) at an ascertained energy is mostly determined by the total
amount of \v{C}erenkov light photons, which the telescope is able
to collect from the $\gamma$-ray showers of that specific primary
energy. In general, the larger the number of photons in the
\v{C}erenkov light flash recorded from an individual atmospheric
$\gamma$-ray shower, the higher the quality of the shower image.

\begin{figure}[!t]
\centering
\includegraphics[width=0.65\linewidth]{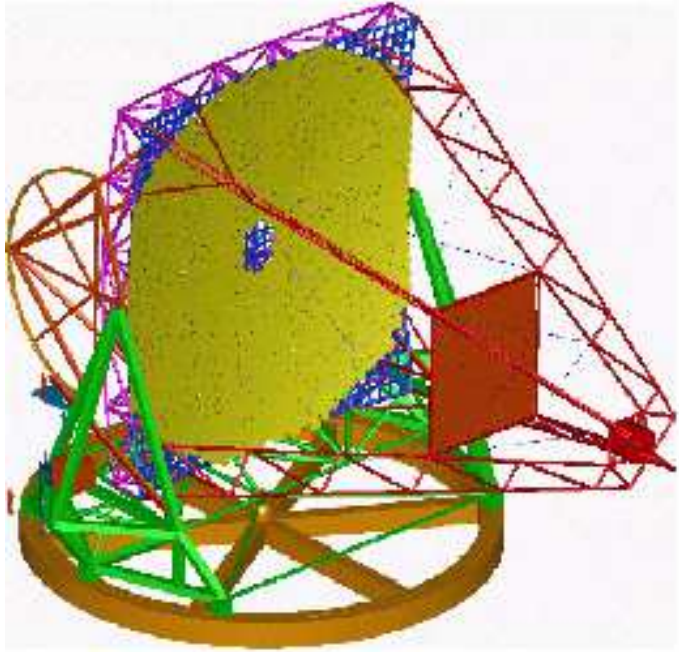}
\caption{The very large air \v{C}erenkov imaging telescope of
600~$\rm m^2$ mirror area and 35~m focal length. Reflector is
segmented into about 850 hexagonal facets of 90~cm width each.
Reflector has a parabolic shape (Courtesy of W. Hofmann).
\label{lay}}
\end{figure}

Three basic parameters account finally for a total number of
\v{C}erenkov light photons registered from the atmospheric shower
by a telescope. First of all is the {\it geometrical size of the
reflector}, $A_o$. It is apparent that the total number of
recorded photons in a \v{C}erenkov light flash scales linearly
with respect to the geometrical area of the telescope's reflector.
A 10~m Whipple \v{C}erenkov telescope has been operating without
major complications for more than 30 years, which has indeed
proven the robustness of a telescope of such size. The H.E.S.S.
instrument comprises four 12~m telescopes. The challenge of
achieving a pointing accuracy of about 20'' and a point spread
function of less than $0.1^\circ$ have been achieved with H.E.S.S.
(Bernloehr et al. 2002). MAGIC group recently started to perform
$\gamma$-ray observations with a 17~m telescope built on La~Palma,
Canary Island. The development and construction of this telescope
has allowed for a positive conclusion to the question of the
technical feasibility of construction of a 30-35~m diameter
\v{C}erenkov telescope (Lorenz, Mirzoyan 2000). The optical
performance of a telescope of such size has been also discussed by
Hofmann (2001) and Akhperjanian, Sahakian (2003). Given all of
those considerations one can indeed envisage that construction of
a 30~m parabolic \v{C}erenkov telescope is indeed practical (see
Figure~\ref{lay}), considering both technical and financial
aspects. Note that telescope reflectors of a diameter far beyond
30~m seem to be extremely costly. Moreover such a reflector will
hardly meet the specifications for the off-axis point spread
function, which is needed for a high-quality imaging of low energy
$\gamma$-ray air showers.

The \v{C}erenkov light photons of an atmospheric shower which hit,
along their transmission path, the telescope's mirrored surface, can
be captured by an advanced imaging camera placed in the focal
plane of the reflector. Using modern PMs (photomultipliers) and
fast electronics it now possible to convert registered photons
into photoelectrons (ph.-e.), and finally digitize the output
signal in a number of ADC (analog-to-digital converter) or FADC
(flash analog-to-digital converter) counts. The overall {\it
efficiency of the photon-to-photoelectron conversion} is another
basic parameter of the telescope design. This efficiency,
$<\epsilon>$, can be calculated as
\begin{equation}
<\epsilon> =(\frac{1}{\lambda_1}-\frac{1}{\lambda_2})^{-1}
\int_{\lambda_1}^{\lambda_2}
r(\lambda)f(\lambda)q(\lambda)p(\lambda)e^{-\sigma(\lambda,d)}
\lambda^{-2} d\lambda
\end{equation}
where $r(\lambda)$ is the mirror reflectivity function, inferred
from a set of measurements performed at different wavelengths;
$f(\lambda)$ is the efficiency of light transmission of Winston
cones, which are placed in front of the camera; $q(\lambda)$ is a
quantum efficiency function of the PMs. The efficiency function,
$p(\lambda)$, accounts for all remaining effects attenuating
light, as for example, the shading the camera by the telescope
masts etc. Usually it totals well below 10\%. The wavelengthes,
$\lambda_1$ and $\lambda_2$, are the boundaries of the wavelength
range of detected \v{C}erenkov photons, which lay within an
optical band essentially between 0.3 and 0.6~$\mu \rm m$. For
conventional telescopes the wavelength-averaged mirror
reflectivity and the transmission of the Winston cones are about
$<r>\sim 0.8$ and $<f>\sim 0.7$, respectively. Note that these
efficiencies are close to the limit of the instrumental capacities
and therefore they may not be substantially improved in the
future. At the same time the quantum efficiency of the
conventional PMs, which is of about $q\, \sim 0.25$ at maximum
value, could be drastically improved for the modern advanced
photo-detectors, e.g. PMs of APD S5345 type provide the quantum
efficiency as high as $q\, \sim 0.8$ (see MAGIC Proposal, 1998).

\begin{table}[t]
\caption{Basic parameters of telescope design for currently
operating telescopes as well as for recently suggested new
projects. QE stands for quantum efficiency. $<A_o>$ is the
effective area of a single telescope (see text). \label{t1}}
\begin{center}
\begin{tabular}{lllll}
Experiment: & Reflector size: & QE: & $<A_o>$ & Altitude of site:
\\ \hline
Whipple     & 10~m & $\sim 0.25$ & 8~$\rm m^2$ & 2.3~km\\
HEGRA       & 5$\times$3.5~m& $\sim 0.25$ & 1~$\rm m^2$ & 2.2~km\\
H.E.S.S. I  & 4$\times$12~m & $\sim 0.25$ & 11~$\rm m^2$ & 1.8~km\\
H.E.S.S. II & 28~m & $\sim 0.25$ & 61~$\rm m^2$ & 1.8~km\\
CANGAROO III& 4$\times$10~m & $\sim 0.25$ & 8~$\rm m^2$ & 160~m \\
VERITAS & 4$\times$12~m & $\sim 0.25$ & 11~$\rm m^2$ & 1.8~km \\
MAGIC I     & 17~m & $\sim 0.25$ & 23~$\rm m^2$ & 2.2~km\\
MAGIC II    & 17~m & $\sim 0.80$ & 73~$\rm m^2$ & 2.2~km\\
ECO 1000    & 36~m & $\sim 0.80$ & 325~$\rm m^2$ & 2.2~km\\
5$@$5        & 5$\times$20~m & $\sim 0.25$ & 31~$\rm m^2$ & 5~km \\
{\sc STEREO ARRAY}& 5$\times$30~m & $\sim 0.25$ & 70~$\rm m^2$ & 1.8~km\\
\hline
\end{tabular}
\end{center}
\vspace*{4mm}
\end{table}

A substantial fraction of \v{C}erenkov light photons emitted in
atmospheric showers attenuates due to absorption and scattering of
light onto the atoms, which Earth's atmosphere is made up of. The
corresponding reduction of the photon flux is given as
$\Phi^{ph}_a \propto e^{-\sigma(\lambda,d)}$, where
$\sigma(\lambda,d)$ is a photon attenuation cross-section at the
wavelength of $\lambda$, and for the effective distance of the
photon emission point to the telescope, $d$. This effective
distance is determined by the development height of an atmospheric
shower, which depends on the primary energy of a shower, as well
as by the {\it altitude of the observational site}, $h_o$. By
setting the detector closer to the shower, which means up on the
high altitudes, one can reduce the effective propagation length of
the photons, $d$, in the atmosphere and correspondingly increase
in the number of \v{C}erenkov light photons arriving onto the
telescope reflector. As Aharonian~et~al (2001) pointed out, the
density of \v{C}erenkov light photons at high altitude, $h_o \geq
5$~km above sea level, can increase, approximately, by a factor of
2 as compared with the corresponding density at conventional
altitudes of $h_o \sim$2~km above sea level. In fact, \v{C}erenkov
light photons are heavily absorbed whilst propagating within a few
kilometers of the very dense atmospheric layer right above the
observational level, which contains noticeable aerosol
contamination. There are a number of high altitude sites in the
world which are accessible for construction of the \v{C}erenkov
telescopes. However, the topology of the \v{C}erenkov light images
recorded at the high altitude site is such, that it makes
implausible to achieve much of a gain in sensitivity from such an
array located at high altitude (Konopelko 2004). Thus we have
considered here a conventional observation level of $h=1.8$~km,
which is close to the optimum one and eventually corresponds to
the current H.E.S.S. site. Note that the H.E.S.S. collaboration
has been funded to construct a 28~m telescope to be added to the
current H.E.S.S. array, which will be place right in the center of
the four telescope system. This telescope might be considered as a
prototype for an array of five 30~m telescopes considered here.

For the conventional PMs the overall photon-to-photoelectron
conversion efficiency is about $<\epsilon>\sim 0.1$, whereas using
enhanced quantum efficiency photo-detectors one can achieve the
conversion efficiency of $<\epsilon>\sim 0.4-0.5$. However, such
photo-detectors are still under development and are currently not
commonly available on the market. Therefore, we considered here a
telescope camera built using conventional PMs and electronics.

Finally the basic parameters of the telescope's design determine
the effective area of the telescope reflector $< A_o >
=<\epsilon>A_o$, which is calculated with respect to a number of
photoelectrons recorded by the camera. For the parameters chosen
for a detector depicted here as {\sc STEREO ARRAY} the
corresponding effective collection area is about $< A_o > \sim \rm
70~m^2$ (see Table~\ref{t1}). Ultimately, in the foreseeable
future the large dish telescopes can be upgraded with cameras made
of advanced photo-detectors, which will push the energy threshold
as well as the sensitivity of this instrumentation to the absolute
physical limit of the imaging atmospheric \v{C}erenkov technique.

\section{Energy Threshold}

The camera pixels of imaging atmospheric \v{C}erenkov telescopes
are able to function under a hard load of night sky background
light, yielding a  photoelectron rate of about 200 MHz per
pixel. The camera trigger, which embraces the PM signals within a
certain (trigger) zone of the camera, is designed so that it
eliminates accidental triggers due to Poisson-like
fluctuations of night sky background light over a large number
of camera pixels.

The optimum and robust trigger criterion allows for a reduction
the rate of the accidental triggers down to a level which is at
least one order of magnitude below the measured cosmic-ray rate.
For a 30 m telescope such criterion requires, for instance, a
simultaneous registration of about 3 pixels with a signal
exceeding a $\sim$6~photoelectron level, which is finally
determined by the telescope and camera design. In this case the
minimal size (total number of photoelectrons in image) of the
triggered events is around 20~photoelectrons. Events of such low
size consist of only a few pixels and do not offer any reliable
measurement of image shape and are in fact almost useless for
stereoscopic shower reconstruction. Despite a few past trials, so
far there is no well established analysis method that been
developed specifically to extract such low size $\gamma$-ray
events. In reality one has to apply a size cut at the level of
40~photoelectrons, which removes all images of low quality from
the data sample on the costs of higher energy threshold. After
applying selection by size the remaining events allow measurement
of the orientation and the angular extensions along the major and
minor axis of an ellipsoid-like image with sufficiently high
accuracy.

\begin{figure}[!t]
\centering
\includegraphics[width=0.7\linewidth]{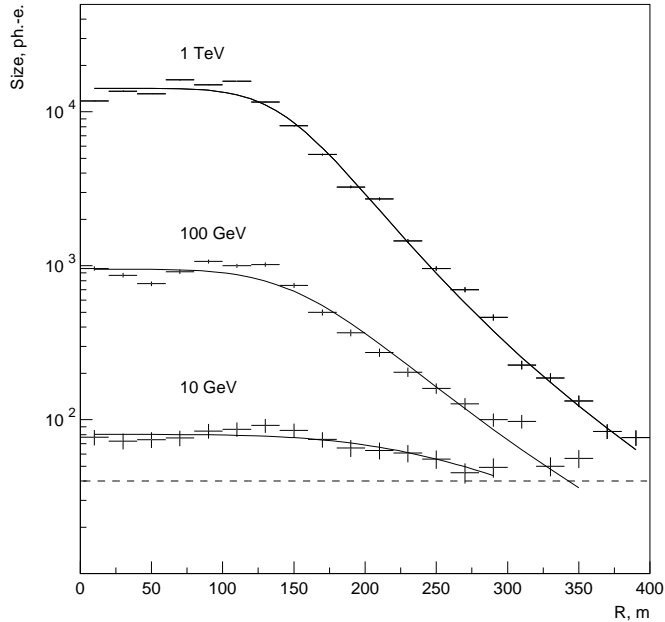}
\vspace*{-2mm} \caption{Lateral distribution of mean image size in
10~GeV, 100~GeV, and 1~TeV $\gamma$-ray atmospheric shower
simulated for a 30~m telescope with a $3^\circ$ camera. Solid
lines show the fit functions given by Eqn.(\ref{fit1}). Dashed
line indicates the minimum acceptable size of 40~photoelectrons.
Parameter {\it Size} corresponds to a total number of
photoelectrons [ph.-e.] in the image. \label{f1}} \vspace*{-1 mm}
\end{figure}

The minimal size of the recorded images is the final constraint on
the actual energy threshold of the telescope. The density of the
\v{C}erenkov light photons emitted in the $\gamma$-ray-induced
atmospheric shower, which was recalculated to a number of
photoelectrons assuming specific telescope design, as a function
of the distance of telescope to the shower axis in observational
plane is shown in Figure~\ref{f1}. For a given reflector size and
efficiency of photon-to-photoelectron conversion one can estimate
corresponding minimal density at the telescope's altitude, which
corresponds to the minimal size cut. The lateral distribution of
the photon density can be well described by a fit like
\begin{equation}
\rho(R)= C(1+(R/R_\circ)^\beta)^{-1}, \label{fit1}
\end{equation}
where the fit parameters $C,\,R_\circ,\,\beta$ for a 100~GeV
$\gamma$-ray shower are 14.0, 264.0, 3.41, respectively. One can
see in Figure~\ref{f1} that a broad plateau expanding up to the
radius of R$\sim$120~m finally develops into an exponential
fall-off\footnote{This fit has a limited accuracy around
R$\sim$120~m. It does not reproduce a very narrow bump, which
occurs exactly at this radius due to high energy shower particles,
but it is sufficiently good for the energy threshold estimate
discussed here.}. This lateral profile of \v{C}erenkov light
density resulted from a specific location of shower maximum in the
atmosphere and a very narrow angle of \v{C}erenkov light emission,
of about 1$^\circ$.

In a zero approximation a total amount of \v{C}erenkov light
photon emitted in a $\gamma$-ray shower is roughly proportional to
the shower primary energy. Therefore, one can scale the density of
\v{C}erenkov light photons or size with respect to shower energy,
$E_\circ$,
\begin{equation}
\rho (E_\circ , R) \simeq \rho (R)(E_\circ / \rm 100\,\, GeV)^{1.2}
. \label{ee2}
\end{equation}
For instance for the density of \v{C}erenkov light photons about
$\rho_\circ\simeq\,0.6\, \rm ph\,m^{-2}$ the corresponding image
size is of $S = < A_o> \rho_\circ \simeq 40$~ph.-e. One can see in
Figure~\ref{f1} that a 10~GeV $\gamma$-ray shower can be
effectively detected by a 30~m telescope. At the same time a 1~TeV
$\gamma$-ray shower at the impact distance of less than 250~m
yields more than $10^3$~ph.-e., which is sufficient to produce a
very high quality image. Note that the showers of a 10~TeV energy
can be in principle detected at a 1~km impact distance, which
results in a very large detection area being available for the
stereo trigger, given a satisfactory large field of view of the
imaging camera.

Despite that various definitions of the energy threshold of an
imaging atmospheric \v{C}erenkov telescope have been discussed in
the literature, at present it is widely accepted as a standard one
the definition of the threshold as the energy, which corresponds
to the position of the maximum in the differential detection rate
of the $\gamma$-rays with the Crab-like energy spectrum in
observations at zenith. The energy threshold of 500~GeV for the
HEGRA system of five imaging atmopsheric \v{C}erenkov telescopes
estimated from the Monte Carlo simulations was fully confirmed by
the direct measurements supported by a number of independent
calibration tools (Konopelko et al. 1999). The same is true for
the H.E.S.S. array of four telescopes in Namibia, which is right
now in a state of routine data taking. The energy threshold of a
100~GeV predicted on the basis of Monte Carlo simulations
(Aharonian et al. 1997) was verified recently using detailed
systematic studies of the $\gamma$-ray fluxes from the Crab Nebula
and other sources (Hinton, 2004). All that strengthen a confidence
of the calculation of the energy threshold for a {\sc STEREO
ARRAY} given in this paper.

\vspace*{-2mm}
\section{Focal Plane Detector}

Modern atmospheric \v{C}erenkov telescopes focus the light from
the atmospheric shower onto a fine-granularity imaging camera. An
angular size of the pixel (PM) and the total field of view are the
basic parameters of camera configuration. Optimum design of a
camera must provide adequate registration of \v{C}erenkov light
images of $\gamma$-ray showers within {\it a priori} defined
dynamic energy range. For the 30~m diameter telescopes the low
energy bound is about 10~GeV. The shower maximum of such low
energy $\gamma$-ray showers is high above the observational level
- $H_{max}\simeq 11$~km. Therefore, the images of these showers
have on average a very small angular size and in addition they are
located very close to the center of the camera's field of view.
For an accurate measurement of angular size of these images one
needs to reduce the angular pixel size down to about 0.07$^\circ$
in order to increase a number of tubes involved in image
parametrization. At the same time small pixel size suppresses the
contamination of the night sky background light accumulated within
a signal readout time window (for detailed discussion see
Konopelko, 2004). Note that further reduction of the pixel size is
inexpedient, because such an extremely small pixel size will be
beyond the reasonable limit given by optical performance of a
$\sim$30~m parabolic dish (Hofmann 2001).

In a toy model one can assume that most of \v{C}erenkov light
photons are coming directly from the shower maximum, $H_{max}$,
which places at the atmospheric depth - $p_{max}\,\rm
[gr\,cm^{-2}]$ - which corresponds to the atmospheric depth of a
maximum number of secondary electrons in a shower. If the
telescope is tracking the source at the center of its field of
view, then the position of the maximal image intensity (centroid)
in the camera focal plane is determined then exclusively by the
height of the shower maximum above the observation level and the
geometrical distance of the shower axis to the telescope. In
approximation A of the classical cascade theory, the shower
maximum has taken place at the depth
\begin{equation}
x = X_\circ ( ln(E_\circ/E_c)-1/2), \,\, \rm gr\, cm^{-2}
\label{xx}
\end{equation}
where $E_\circ$ is a primary energy of a $\gamma$-ray shower;
$E_c$ is the so-called critical energy ($E_c\simeq 80$~MeV), and
$X_\circ$ is a radiation length in atmosphere ($X_\circ =
37.1\,\,\rm gr\,cm^{-2}$). For a realistic atmospheric model the
relation between the height, $h$~[m], and the atmospheric depth,
$x\,\, \rm [gr\,cm^{-2}]$, can be given as
\begin{equation}
h=(6740+2.5\, x)ln(1030/x). \label{model}
\end{equation}
Finally the angular shift of the centroid can be calculated as
\begin{equation}
\Theta = \arctan [R\,(h-h_\circ)^{-1}], \label{tt}
\end{equation}
where $h_\circ$ is a height of the observational level above sea
level (hereafter  $h_\circ \simeq 1.8$~km). For instance the shift
of image centroid of a 10~GeV $\gamma$-ray shower at 200~m from
the shower axis is about 1.25$^\circ$. Note that such toy model is
accurate enough in most cases but it introduces a
systematic error due to difference between the actual shower
maximum position and the shower maximum as seen in the
\v{C}erenkov light, which is partially absorbed whilst propagating
into the atmosphere.

\begin{figure}[t]
\begin{center}
\includegraphics[width=0.7\linewidth]{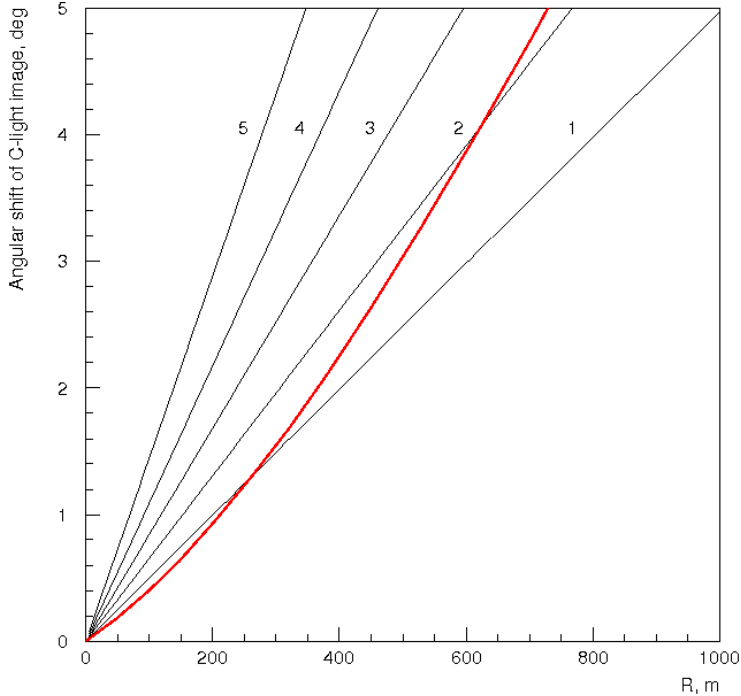}
\caption{An angular shift of image centroid from the center of the
camera focal plane for $\gamma$-ray showers of a different primary
energy. Curves 1, 2, 3, 4, 5 correspond to the showers of 10,
$10^2$, $10^3$, $10^4$, and $10^5$~GeV, respectively. Bold curve
limits the range of impact distances, which are allowed by the
requirement of a minimal \v{C}erenkov light density for shower
registration ($\rho = \rm 0.6~ph\,m^{-2}$). No restriction on
angular acceptance was applied here. \label{dist}}
\end{center}
\end{figure}

For a fixed primary energy Eqn. (\ref{ee2}) can be inverted and
the maximum radius, providing sufficient number of \v{C}erenkov
light photons for shower registration, can be calculated. Using
Eqn. (\ref{tt}) one can estimate the corresponding angular shift
of the image centroid in the camera focal plane. For example for a
1~TeV $\gamma$-ray shower the image displacement is about
1$^\circ$ at a 100~m impact distance\footnote{Note that this case
deals with a telescope pointing directly towards the source.
However, the discussion given here also generally applies to other
telescope pointings.}. Bold curve in Figure~\ref{dist} indicates
the allowed range of the angular shifts for $\gamma$-ray showers
of different primary energies. One can see in Figure~\ref{dist}
that for a 10~GeV energy $\gamma$-ray the image has a relatively
small displacement from the center of field of view, which is
below $1.5^\circ$. The $\gamma$-ray showers of higher energies can
trigger the detector at much large impacts if the angular size of
the camera is sufficiently wide to accommodate these events.

The point spread function of a 30~m parabolic reflector degrades
relatively fast at $\geq 1.5^\circ$ off the telescope
optical axis. For example for a 17~m MAGIC telescope approximately
50\% of all reflected light is concentrated within a circle of
$0.1^\circ$ for $1^\circ$ off-axis observations (see MAGIC
proposal). For a 30~m telescope assuming a realistic focal length
(F/D=1.25) the point spread function will finally limit the
angular range of efficient imaging by approximately $\sim
1.5-2.0^\circ$ off the optical axis. For larger declinations of
incoming photons the point spread function becomes substantially
wider than the actual angular size of the image.

The high energy $\gamma$-ray showers propagate deep into the
atmosphere. Therefore their shower maximum occurs at relatively
small heights above the observational level. If the telescope is
pointing directly towards the source then the images of triggered
high energy events will be concentrated in the camera focal plane
at rather large angular distances from its center. An imaging camera
with a limited angular size (field of view) of $3.0^\circ$ diameter
will be able to detect only \v{C}erenkov light photons coming from
the upper layers of the atmosphere, i.e. from the forefront of the
shower development. This reduction of detection efficiency for
very high energy $\gamma$-rays ($E_\circ\geq1$~TeV), due to a
camera's limited  field of view, becomes even more severe for the
telescopes deployed at high altitudes. In this case the shower
maximum can be only observed at very large angles with respect to
the telescope optical axis.

\begin{figure}[t]
\begin{center}
\includegraphics[width=0.7\linewidth]{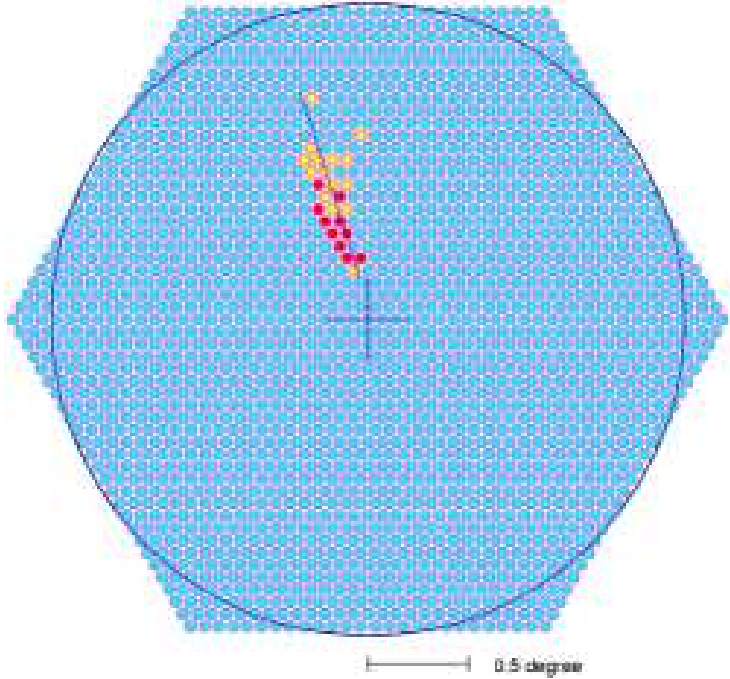}
\caption{PMs pattern in a 1951 pixel camera. Pixel size is
0.07$^\circ$. Superimposed is the \v{C}erenkov light image from a
30~GeV $\gamma$-ray shower registered at the impact distance of
50~m from the shower axis. Solid line indicates the reconstructed
orientation of the major axis of the image. Pixels marked in grey
have amplitudes in a range of 4-10~ph.-e. Pixels with amplitude
above 10~ph.-e. are shown in dark grey. The pixels are shown here
as circles, whereas in the simulations they were in fact
considered to be the hexagons. The solid black circle has a radius
of 1.5$^\circ$, which is the effective angular size of the camera.
\label{pixels}}
\end{center}
\end{figure}

There is always a trade-off between the angular size of the pixel
and the total number of pixels in the camera, which are needed to
cover a certain field of view. Cameras of a very small pixel size
account for an enormous number of pixels/channels, which is,
firstly, very expensive and secondly, very difficult in a long run
exploitation. The pixel size of $0.07^\circ$ for the field of view
of 3$^\circ$ diameter is in fact a reasonable choice. It results
in 1951 pixels, which is only by factor of two larger than that
for the currently operating H.E.S.S. I cameras. An artist's view
of a possible camera design is shown in Figure~\ref{pixels}.

We decided to use the conventional ADCs for the signal
digitization. At present the question of the advantageous performance
of the FADCs for  data acquisition and even for $\gamma$-ray
imaging is still under general discussion and  needs
further verification.

\section{System Layout}

The HEGRA ({\it High Energy Gamma Ray Astronomy}) experiment has
proven for the first time numerous advantages of the stereoscopic
observations of $\gamma$ rays above 500~GeV from the ground.
H.E.S.S. ({\it High Energy Stereoscopic System}) was a natural
extension of the HEGRA approach into lower energy range starting
from 100~GeV. Utilizing a stereoscopic approach H.E.S.S. achieved
the sensitivity at the level of $\sim 1$\% of the Crab Nebula
(standard candle) flux, which is so far unique in the field of
very high energy $\gamma$-ray astronomy. Further reduction in
energy thresholds down to $\sim$10~GeV also envisages making use
of the stereoscopic approach. Large fluctuations in low energy
showers noticeably degrade the quality of recorded \v{C}erenkov
light images. However, this can be substantially remedied using
stereoscopic observations. To perform stereoscopic observations at
low energies one needs apparently to have at least two 30~m
imaging \v{C}erenkov telescopes operating synchronically. Such a
system of two telescopes might be considered as a basic
stereoscopic system for observations of low energy $\gamma$ rays.
Here we present the results of simulations for a 30~m single
stand-alone telescope, as well as for a pair of such telescopes,
and finally for a {\sc Stereo Array} of five 30~m telescopes,
which might be considered as a prototype for a future major
facility in the field of very high energy $\gamma$-ray astronomy.
The layout of the five telescopes is similar to that of the HEGRA
layout but with a spacial separation between the telescopes of
100~m. A 100~m separation corresponds to a geometrical size of a
\v{C}erenkov light pool on the ground, which is in fact of about
100~m in radius. Despite the fact that a search for an optimum
separation between two 30~m telescopes is an interesting issue,
which needs perhaps special dedicated investigations, and might
slightly influence the final sensitivity value, it is out of the
frame of present paper, but it will be discussed in a separate
forthcoming paper.

\begin{table}[t]
\caption{Basic parameters of the simulational setup.\label{t2}}
\begin{center}
\begin{tabular}{ll}\hline
Altitude:    & 1.8~km above sea level \\
Atmosphere:  & Tropical \\
Reflector size: & 30 m\\
Reflector design: & parabolic (F/D=1.25) \\
Focal length: & 37.5 m \\
Number of telescope: & 5 \\
Distance between telescopes: & 100~m \\
Number of camera pixels: & 1951 \\
Pixel size: & 0.07$^\circ$ \\
Photon-to-photoelectron efficiency:  & $\sim$0.1\\
Trigger:    & Signal in each of 3 PMs exceeds 6 ph.-e. \\
Tail cut:   & 3/5 ph.-e. \\ \hline
\end{tabular}
\end{center}
\end{table}

\section{Simulations}

Atmospheric showers induced by $\gamma$-rays, electrons, and
protons have been simulated using a numerical code described by
Konopelko, Plyasheshnikov (2000). The primary energy of simulated
showers was sampled as uniformly distributed within each of 24
bins, which were chosen within the energy range starting from
1~GeV and extending up to 10~TeV. Five 30~m \v{C}erenkov
telescopes were included in the simulational setup. The maximum
impact distance of the shower axis to the center of the array was
1000~m. Basic parameters of the simulational setup are summarized
in Table~\ref{t2}. The detailed procedure simulating the camera
response was applied for the generated events. This procedure
accounts for all possible losses of the Cherenkov light from the
air shower on the way from the mirror reflecting to the camera
response. The list of the effects which are important in this
respect contains: {\it (i)} mirror reflecting sampled by the light
{\it raytracing} technique or using the measured functions of the
light spot distortion in the camera focal plane; {\it (ii)} PMT's
funnel acceptance; {\it (iii)} photon-to-photoelectron conversion
inside the PMT's taking into account the measured {\it single
photoelectron spectrum}. Further details of the simulational
procedure can be found in Konopelko et al. (1999).

\section{Energy Spectra of Primary Particles}
\label{spectra}

In general a \v{C}erenkov telescope counts events, which are
atmospheric showers initiated by various particles of primary
cosmic rays. Their major components are electrons,
protons, and nuclei. The estimate for detector sensitivity is
usually computed with respect to the well-established DC
$\gamma$-ray flux from the Crab Nebula. The spectra of
$\gamma$-rays and different background components measured in
relevant energy range, which have been used in the sensitivity
estimate given here, are summarized below.

\subsection{Gamma-rays}

The DC flux of VHE $\gamma$-rays from the Crab Nebula is generally
accepted as a standard flux unit in $\gamma$-ray astronomy. The
spectrum of VHE $\gamma$-rays from the Crab Nebula was measured
with a number of ground-based \v{C}erenkov light detectors in
energy ranges, starting from 50~GeV and extending up to 50~TeV
(for review see e.g. Aharonian et al. 2000). The energy spectra
and absolute fluxes measured by different instruments over
different energy intervals are generally compatible. Therefore it
is reasonable to use well-established spectrum of VHE
$\gamma$-rays from the Crab Nebula for an estimate of the
sensitivity of future \v{C}erenkov light instruments and finally
give a sensitivity merit expressed in Crab flux units.

The Crab Nebula spectrum measured with the HEGRA system of 5
imaging atmospheric \v{C}erenkov telescopes can be
approximated by a power-law with a logarithmic steepening
(Aharonian~et~al. 2000):
\begin{eqnarray}
dF_{\gamma}/dE = \rm (2.67 \pm 0.5) \times 10^{-11}
(E/1\,\,TeV)^{-2.47
\pm 0.15 - (0.11 \pm 0.10) log(E)}, \nonumber \\
\rm  cm^{-2} s^{-1} TeV^{-1} \label{crab_spectrum}
\end{eqnarray}
The best fit indicates a slight flattening of the spectrum at the low
energies, as is predicted by theories which assume an Inverse
Compton (IC) origin of VHE $\gamma$ rays from the Crab Nebula
(e.g. see de~Jager~et~al. (1996); Atoyan \& Aharonian (1999)).

An extension of the HEGRA Crab Nebula spectrum to low energies,
achieved with HEGRA by using a specific observational technique
(Lucarelli~et~al. (2003)), is fully consistent with the recent
measurements carried out with the Solar type detectors. Note that
recent measurements of the Crab Nebula spectra in the energy range
above 10~TeV (Aharonian et al. 2004) with HEGRA, and  the
preliminary data obtained with H.E.S.S. (Masterson et al. 2004)
are in agreement. In fact the best fit of the HEGRA data, given by
Eqn.(\ref{crab_spectrum}), describes rather well all various
spectral and flux measurements obtained so far. This fit was used
here in computing the $\gamma$-ray detection rates. Note that the
HEGRA and EGRET spectra of the DC $\gamma$-ray flux from the Crab
Nebula are consistent within statistical and systematic errors
given for both experiments.

\subsection{Electrons}

For the ground-based atmospheric \v{C}erenkov telescopes, which
are approaching the energy range of $\sim$10~GeV, the electron
component of primary cosmic rays becomes a dominant background and
in fact is an issue of major concern. Around 10~GeV the
intensity of the electron component is about $\sim$1\% of the
corresponding proton intensity. The \v{C}erenkov light images
generated by electron-induced atmospheric showers are almost
indistinguishable in shape from the images of the $\gamma$-ray
showers. Moreover the angular imaging resolution of  \v{C}erenkov
telescopes substantially worsens at such low energies, and indeed
 does not allow for very strong suppression of an isotopic flux of
cosmic electrons. Finally, the performance of imaging atmospheric
\v{C}erenkov telescopes in the energy range around 10~GeV
relies heavily upon the residual trigger rate of electron-induced showers
obtained after application of the analysis cuts.

A number of various measurements of the energy spectrum of cosmic
electrons have been performed in past (see Du~Vernois~et~al.
2001). Despite that the existing data exhibit sizeable statistical
and systematic uncertainties above 10~GeV the electron spectrum is
a power-law with the spectral index of about 3.0. Over a broad
energy range from 1~GeV up to 100~GeV the electron
spectrum\footnote{Hereafter we consider only the combined spectrum
of $e^+$ and $e^-$.} can be fitted as
\begin{eqnarray}
F_e/dE = \rm 1.2 \times 10^{-3} E^{-1}(1+(E/5\,\,GeV)^{2.3})^{-1},
\,\,\, \rm cm^{-2}sr^{-1}s^{-1}GeV^{-1}\,\, \label{ee}
\end{eqnarray}
Such a steep energy spectrum makes it almost impossible to detect low
fluxes of cosmic-ray electrons at high energies with currently
operating IACTs, whereas absolute electron flux at energies of
only a few GeV is relatively high. At low energies the flux of electrons
becomes comparable with the flux of cosmic-ray protons after
applying analysis cuts and it should be taken into account in
total background estimate. Here we computed the electron rates
using the spectral fit given by Eqn.(\ref{ee}).

\subsection{Cosmic-Ray Protons \& Nuclei}

The fluxes and spectra of primary cosmic-ray protons and helium
nuclei can be thoroughly measured in the energy range of 1 to
200~GeV with the balloon born experiments. These spectra have been
recently upgraded using the BESS spectrometer with the accuracies of
$\sim 5$\% for protons and $\sim 10$\% for helium (Sanuki et al.
2000)\footnote{In this paper a most complete compilation of modern
data on differential proton and helium spectra is also given.}.

In the energy range of 1 to 100~GeV the 1998 BESS data on proton
spectrum can be reproduced using the following fit
\begin{eqnarray}
dF_{\sc CR}/dE = \rm 0.3907\,E^{0.3016}(1+(E/1.776~GeV))^{-3.141},
\nonumber \\ \rm cm^{-2}s^{-1}str^{-1}GeV^{-1} \label{cr1}
\end{eqnarray}
Above 30~GeV the proton spectrum is a straight power law
\begin{equation}
dF_{\sc CR}/dE= \rm 9.6\times 10^{-9}(E/10^3~GeV)^{-2.7},\,\,
 \rm cm^{-2}s^{-1}str^{-1}GeV^{-1}. \label{cr2}
\end{equation}
Note that the slope of a power law spectrum is consistent with the
spectra given by Wiebel et al. (1994) and Simpson et al. (1983),
which are $dF_{\sc CR}/dE \propto E^{-2.75}$ and $dF_{\sc CR}/dE
\propto E^{-2.66}$, respectively, whereas the absolute fluxes
differ by 20-50\%. Since at present BESS measurements offer the
best quality data, the proton spectrum described by
Eqns.(\ref{cr1},\ref{cr2}) was used here in calculations of the
cosmic-ray detection rates.

It is known that Helium nuclei in primary cosmic rays contribute
up to $\sim$30\% to the total raw event rate of atmospheric
\v{C}erenkov telescopes (see e.g. Konopelko et al. 1999). The
contribution of all other cosmic-ray nuclei is very low and almost
negligible. That was proven here by specific test simulations. In
a good approximation one can scale the event rate calculated for
the proton cosmic ray flux by a factor of 1.5 in order to get a
total cosmic ray detection rate.

\section{Performance of {\sc STEREO ARRAY}}

An estimate of the detector performance is inferred from the
detection rates of $\gamma$-ray and background events, calculated
after application of the analysis cuts. For a given flux of
cosmic-ray particles of a specific type the energy dependent
function of correspondent detector response, i.e. collection area,
basically determines relevant detection rate. Collection areas for
$\gamma$-ray- and cosmic-ray- induced showers can be derived from
the detailed Monte Carlo simulations tracing both the shower
development in the atmosphere and the detector performance. Below we
summarize the basic results on collection areas of $\gamma$ rays
and background events of a different kind.

\subsection{Collection Areas}

In the simulations the position of a shower axis in the
observational plane was sampled as uniformly randomized over an
area limited by 1~km radius from the center of the array. By
counting the number of triggered showers of a specific primary energy
one can calculate a so-called trigger probability, which then can
be easily converted into corresponding collection area. Events
were simulated within the energy range from 1~GeV up to 10~TeV.

\begin{figure}[!t]
\centering
\includegraphics[width=0.7\linewidth]{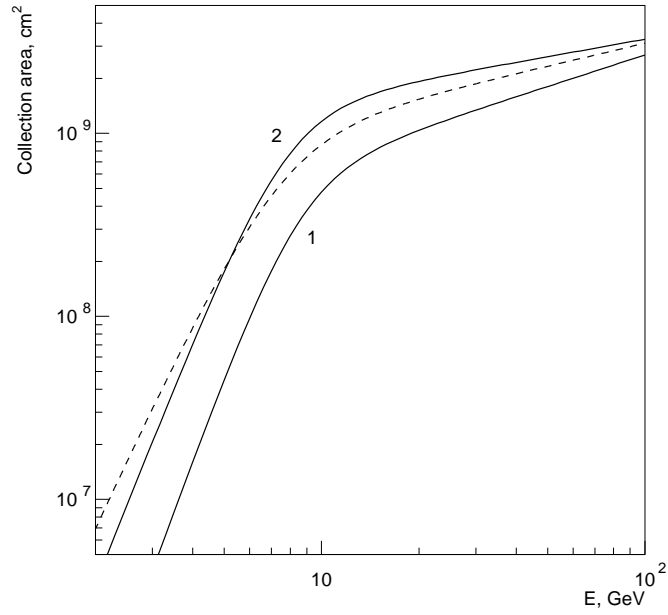}
\includegraphics[width=0.7\linewidth]{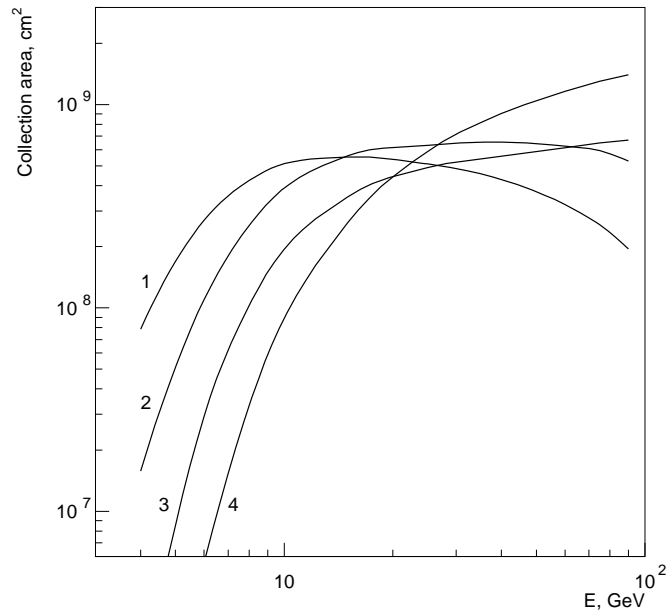}
\caption{The collection areas of $\gamma$-ray showers for a single
stand-alone 30~m IACT (dashed curve), for a system of two 30~m
IACTs with a 100~m separation (curve 1) and for a system of five
30~m IACTs for 2-fold telescopes coincidences (upper panel). The
collection areas for a system of five 30~m IACTs for multiplicity
of 2, 3, 4, 5 telescopes (curves 1,2,3,4, respectively, lower
panel).\label{areas}}
\end{figure}

The energy dependence of the collection area basically reflects the
profile of the \v{C}erenkov light pool on the ground. For
$\gamma$-ray showers it has a high density plateau expanding up to
120~m distance from the shower axis, which then develops into an
exponential fall-off. One can see in Figure~\ref{areas} that a
steep rise of collection area at low energies (E$\leq$50~GeV)
converges into a rather slow logarithmic growth at high energies.
For $\gamma$-ray showers the characteristic energy at which the
energy profile of collection area breaks roughly corresponds to
the energy threshold of the instrument. For proton showers the
\v{C}erenkov light pool does not have such a prominent
``shoulder'' at 120~m but it yields rather smoothly decreasing
photon density with enlargement of the distance to the shower
core.

The collection areas of $\gamma$-ray atmospheric showers can be
consummately reproduced by a four parameter fit:
\begin{equation}
S_{\gamma} = a_1E^{a_2}(1+(E/a_3)^{a_4})^{-1},\,\, \rm cm^2,
 \label{parm}
\end{equation}
The parameters of the fit for different detector configurations
are given in Table~\ref{parm1}.

\begin{table}[t]
\caption{Parameters of a fit of collection area given by
Eqn.(\ref{parm}).\hfill\break ($^*$ The system trigger required at
least 2-fold coincidence).}
\begin{center}
\begin{tabular}{llllll}\hline
Configuration: & $a_i$ & i=1 & 2 & 3 & 4 \\ \hline Single
Stand-Alone
Telescope & & $5.03\times 10^5$ & 3.80 & 7.50 & 3.38 \\
Stereo: Two Telescopes & & $1.98\times 10^4$ & 4.86 & 8.49 & 4.29 \\
Stereo: Five Telescopes$^*$ & & $1.58\times 10^5$ & 4.45 & 7.77 &
4.13 \\ \hline \label{parm1}
\end{tabular}
\end{center}
\end{table}

The proton-induced showers are isotropically distributed over the
incidence angle with respect to the telescope optical axis. Thus
in this case the collection area has to be computed by averaging
over a solid angle, limited by an opening angle, which exceeds by
roughly $1^\circ$ the opening angle of the camera's field of view.

Interestingly, the second parameter of the fit, $a_3$
(Eqn.(\ref{parm})), roughly corresponds to the energy threshold of
the instrument with respect to the corresponding type of
atmospheric showers, which is about 8~GeV for $\gamma$-ray showers
and 40~GeV for proton-induced showers.

The trigger efficiency for atmospheric showers of some specific
energy is finally determined by a total amount of \v{C}erenkov
light emitted in the shower of such an energy (see discussion above).
For low energy proton-induced showers ($E<$100~GeV) less than 30\%
of their primary energy can be transferred into electromagnetic
component, which is finally responsible for the \v{C}erenkov light
emission. That explains why the energy thresholds for the
$\gamma$-ray- and proton-induced atmospheric showers differ greatly.

The development of electron-induced showers in the atmosphere is almost
identical to the $\gamma$-ray-induced showers. A slight difference
in a height of shower maximum for the two cases does not noticeably
effect the lateral and angular parameters of \v{C}erenkov light
pool at the observational level.

In observations of a point-like $\gamma$-ray source the
isotropically distributed background events of cosmic-ray protons
and electrons are registered within a solid angle, which is
limited by actual angular resolution of the instrument. The
angular resolution strongly depends on primary energy of
$\gamma$-ray shower (see below). For {\sc STEREO ARRAY} even at
very low energies, i.e. around the instrument's energy threshold
of $\sim$10~GeV, the angular resolution is still rather good,
$\Theta \simeq 0.3^\circ$. It is substantially less than the
angular size of the camera, which is of $3^\circ$ diameter.
Deviation of the incident angles of electron-induced showers
within $0.3^\circ$ from the telescope optical axis does not affect
the physics of a shower development in the atmosphere, namely the
lateral and longitudinal profiles of secondary shower particles
etc. Therefore, it is possible to use the collection areas
calculated for the $\gamma$-ray-induced showers in calculating the
collection areas of electron-induced showers. Additional test
simulations of some electron-induced atmospheric showers have
proven such approximation to be valid.

In Figure~\ref{areas} the collection areas for a system of five
telescopes for different trigger multiplicities are also shown. It
is important to note that the $\gamma$-ray events of energy below
10~GeV trigger at most only two of the telescopes. The 3-fold coincidences
are generated by $\gamma$ rays of noticeably higher energy.
Finally, the collection area of 5-fold coincidence events
dominates at the energy above 30~GeV. In fact, the low energy
$\gamma$-ray showers can trigger a telescope at quite small impact
distances to the shower axis, which explains the energy dependence
of the collection areas for different trigger multiplicities.

\begin{figure}[!t]
\centering
\includegraphics[width=0.7\linewidth]{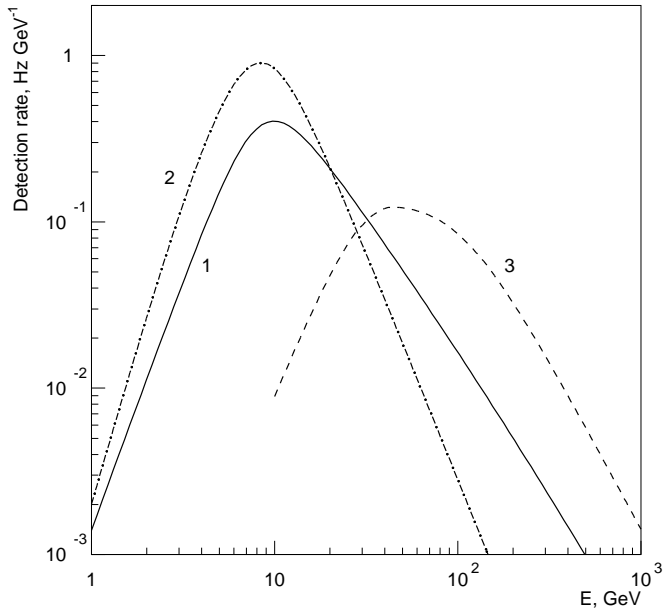}
\caption{Differential detection rates of atmospheric showers
initiated by $\gamma$ rays (curve 1), electrons (curve 2), and
cosmic rays (curve 3). The calculations have been done using the
fluxes given in Section~\ref{spectra}. \label{rates1}}
\end{figure}

\subsection{Detection rates}

Differential detection rates for different types of atmospheric
showers registered with {\sc STEREO ARRAY} can be calculated as
\begin{equation}
dR_{(\gamma,e,CR)}/dE = (dF_{(\gamma,e,CR)}/dE)
S_{(\gamma,CR,e)}\,\,\, \rm Hz\,GeV^{-1} \label{rates}
\end{equation}
The position of a peak in the differential detection rate of
$\gamma$-ray atmospheric showers, which is calculated assuming the
Crab Nebula energy spectrum, defines the effective energy
threshold of the instrument (see Figure~\ref{rates1}). Note that
it is, in fact, in agreement with the value of the $a_3$-parameter
of a fit of the collection area (see Table~\ref{parm1}). One can
see in Figure~\ref{rates1} that the electron-induced showers
dominate at low energies, whereas their contribution is negligible
at high energies due to the very steep energy spectrum of cosmic
ray electrons.

The differential detection rate of the proton-induced atmospheric
showers peaks at substantially higher energy, within 30-90~GeV.
The proton-induced atmospheric shower of that energy yields
approximately the same amount of \v{C}erenkov light as the
$\gamma$-ray showers of about 10~GeV.

The integral raw background rate for a system of two 30~m
telescopes is expected to be about 1~kHz. At the same time the raw
detection rates for a single 30~m telescope, as well as for a
system of five such telescopes, are expected to be of 1.7 and 3.2
kHz, respectively. With up-to-date electronics one can achieve a
sufficiently low dead-time-of-event processing, and it is
generally possible to maintain such high event rate with an
advanced imaging camera. However the discussion of possible
solutions to the problems of trigger electronics and data acquisition
is beyond the framework of present studies.

\subsection{Analysis}

The raw observational data taken with the ground-based atmospheric
\v{C}erenkov telescopes are entirely background dominated. The
detection rate of the background cosmic rays and electrons usually
exceeds by three orders of magnitude the expected rate from a Crab
like $\gamma$-ray source. A detailed comparative analysis of
space, angular, and temporal characteristics of 10~GeV and 100~GeV
$\gamma$-ray showers is given in Konopelko (1997). The images of
the sub-100~GeV $\gamma$-ray showers are indeed very irregular in
shape and are strongly affected by high fluctuations in air shower
development. However they still carry information about shower
orientation as well as its lateral, and longitudinal spread. In
order to extract the tiny fraction of the $\gamma$-ray events from
the enormous amount of background contaminations one has to apply
specific analysis cuts. The set of the analysis cuts (selecting
criteria) usually consists of the selections on the {\it image
orientation} and {\it shape}. Two telescopes make it possible to
perform full-scale stereo reconstruction of the shower direction
in space and on the ground.

\begin{figure}[!t]
\centering
\includegraphics[width=0.7\linewidth]{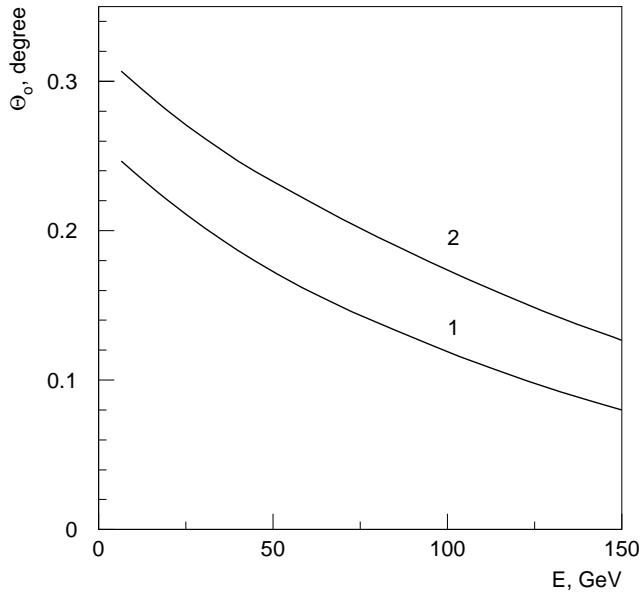}
\caption{The angular resolution of the $\gamma$-ray showers
calculated for a system of two (curve 1) as well as three (curve
2) 30~m atmospheric \v{C}erenkov light telescopes. \label{res}}
\end{figure}

\subsubsection{Orientation:}
The deflection of the reconstructed arrival direction of the
$\gamma$-ray showers from the nominal source position is used for
their orientational selection. The better the angular resolution
of the instruments, the lower the content of residual background
events. The angular resolution (angular radius limiting the area
around the source position accounting for 63\% of all simulated
showers) as a function of primary energy of the
$\gamma$-ray-induced atmospheric shower is shown in
Figure~\ref{res}. At low energies the \v{C}erenkov light images
contain less light and are strongly affected by shower
fluctuations. Therefore the results on reconstruction of the image
orientation as well as the shower arrival direction substantially
worsen. For the angular size of the camera of $\Theta_o \simeq
1.5^\circ$ in radius the cosmic ray rejection is about $\eta \sim
0.04$. One can see in Figure~\ref{res} that the angular resolution
achieved with three images is 30\% better than in the case of only
two images. It allows the reduction of the cosmic ray background
contamination by a factor of 1.7. However, as already mentioned
above, 3-fold triggers substantially decrease the rate of $\gamma$
rays of energy close to the threshold. That is why to search for
$\gamma$-rays at the energy of 10~GeV and below the 2-fold
coincidence mode seems to be more advantageous.

\subsubsection{Image shape:}
\begin{figure}[!t]
\centering
\includegraphics[width=0.6\linewidth]{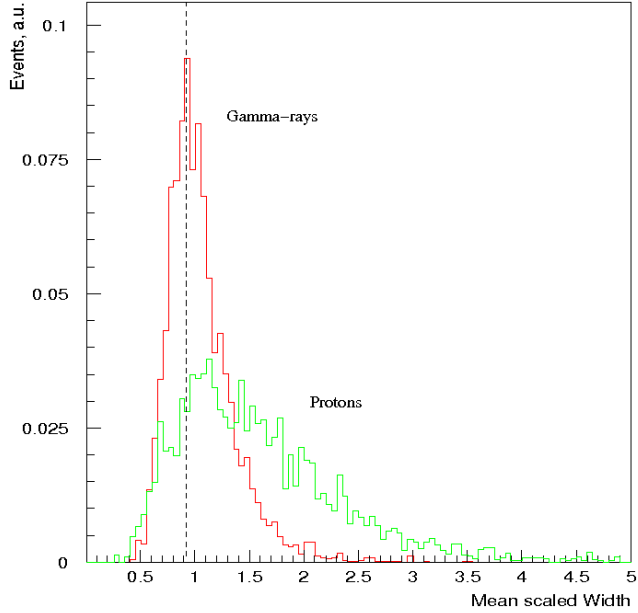}
\caption{Distribution of the {\it mean scaled Width} for the
$\gamma$-ray as well as for the cosmic ray events. By definition
the $\gamma$-ray distribution peaks at 1.0. Mean scaled Width
parameter is calculated as $<w> = \frac{1}{n} \sum_{i=1}^n
W_i/W(R,Size)$, where $W_i,\,i=1,2$ is an actual {\it
Width}-parameter for of $n$ images, $W(R,Size)$ is the mean {\it
Width}-parameter at the corresponding reconstructed shower impact
distance $R$ (m) and image {\it Size} (ph.-e.). Simulated events
are weighted according to the spectra given in
Section~\ref{spectra}. \label{msw}}
\end{figure}

Exploitation of the HEGRA system of 5 imaging atmospheric
\v{C}erenkov telescopes with H.E.S.S. allowed us to conclude that
the most effective parameter of cosmic ray discrimination by the
image shape is a parameter of {\it mean scaled Width (m.s.w.)}. At
least with two telescopes operating in a stereoscopic mode one can
reconstruct the impact distance of the shower axis from each of
the telescopes, and scale the actual transverse angular size of
the image with the total amount of light in the image
($Size$-parameter) at that impact distance. The distributions of
the parameter of mean scaled Width for the simulated $\gamma$-ray
and proton showers are shown in Figure~\ref{msw}. Despite a
significant overlap between two distributions one can reduce the
content of the proton showers by factor of $\sim 10$ (i.e. the
"after cuts" acceptance of cosmic ray showers is
$\kappa_{CR}=0.08$) applying mean scaled Width cut of 0.91. This
cut keeps 35\% of all registered $\gamma$-ray showers. The
corresponding quality-factor is Q-factor=$\kappa_{\gamma}\cdot
(\kappa_{CR})^{-1/2}\simeq 1.2$. High fluctuations in the image
$Size$ of low energy events cause such rather modest rejection
power. Simulations have shown that an application of any other
additional parameter of image shape does not improve the resultant
quality factor.

In observations of a point like $\gamma$-ray source the recorded
images of the $\gamma$-ray-induced air showers are located in the
camera focal plane mostly within the angular annulus limited by
inner and outer radii of 0.3$^\circ$ and 1.2$^\circ$ from the
camera center, respectively\footnote{Note that due to the high
rate of the low energy events they are basically dominant in the
entire sample of registered $\gamma$-rays.}. Thus before applying
the orientation and shape cuts one can noticeably enhance the
content of the $\gamma$-ray showers by using the $Distance$ cut
(angular distance of the image centroid to the camera center).
Such selection criterion allows for the reduction of the rate of
the background events by a factor of 2.

Since the shape of the images generated by the electron- and
$\gamma$-ray-induced atmospheric showers are almost identical, one
can in both cases use the acceptances of the analysis cuts derived
for the $\gamma$-ray-induced showers.

The analysis of data taken with a single stand-alone telescope at
shower energies around 10~GeV is not straight-forward. Note that
there were a number of trials using advanced, sometimes very
sophisticated, analysis methods to improve the rejection of the
background and consequently to enhance the $\gamma$-ray abundance
in data taken with a single telescope. However none of them
demonstrated drastically improved performance as compared with the
standard multi-parametric approach ("super-cuts"). Recent studies
on the multi-variate analysis in application to the data simulated
for a 30~m telescope have shown that a realistic quality-factor
for such low energy events is of about $Q\simeq 3$, which
corresponds to the $\gamma$-ray and cosmic ray acceptance of
$\kappa_\gamma \simeq 0.3$ and $\kappa_{CR} \simeq 0.01$,
respectively (Konopelko et al. 2005). This was achieved using
Bayesian decision rules and a non-parametric estimate of the
multi-parameter density.

The analysis cuts for a single telescope as well as for a system
of telescopes were selected here primarily on the basis of equal
$\gamma$-ray acceptance across the entire dynamic energy range.
Therefore these cuts do not affect noticeably the position of the
peak in the differential detection rate and the estimated value of
the energy threshold.

\subsection{Sensitivity}

The sensitivity of the \v{C}erenkov telescope in observations of
the point-like $\gamma$-ray sources can be defined in different
ways. One approach, which is widely used, is to calculate the
minimum flux of the $\gamma$ rays, which can be detectable after
50~hrs of observations at the significance level above 5$\sigma$
($\sigma$ is one standard deviation of a number of background
events). The total number of $\gamma$-ray, proton, and electron
showers after applying analysis cuts, which remains in the data
sample of observational time $t$, can be calculated as
\begin{equation}
N_{(\gamma,p,e)} = t \int_{E_0}^{E_\infty}
(dJ_{(\gamma,p,e)}/dE)\tilde S_{(\gamma,p,e)}(E)
\kappa_{(\gamma,p,e)}^{d}\kappa_{(\gamma,p,e)}^{s}dE \label{num}
\end{equation}
where $E_0$ is the selected\footnote{Here the selection of the
energy threshold is defined by the physics of the source, which
finally defines the favorable energy range.} energy threshold
energy. $E_\infty$ is the upper energy boundary, which is chosen
well above the actual threshold of the detector ($E_\infty$ is
usually the maximal energy of the simulated showers, and
$E_\infty\sim 10^3 E_{th}$) allowing accurate calculation of the
detection rates. The detection areas of proton and electron
showers are collection areas averaged over the solid angle
$\Omega_o \simeq 2 \pi \theta_{max}, \theta_{max} = 4^\circ$. Thus
the collection areas used in Eqn. (\ref{num}) are $\tilde S_\gamma
= S_\gamma$, and $\tilde S_{(p,e)}=<S_{(p,e)}> \Omega_o$.
$\kappa_{(\gamma,p,e)}^{(d,s)}$ are the acceptances of the
$\gamma$-ray-, proton-, and electron-induced showers after
applying directional and shape cuts, respectively. The integral
detection rates for the $\gamma$-ray, electron, and proton showers
are summarized in Table~\ref{flux}.

The corresponding signal-to-noise ratio after 50~hrs exposure can
be calculated as
\begin{equation} (S/N) = N_\gamma\cdot (N_p + N_e)^{-1/2}
\end{equation}
In order to derive the $\gamma$-ray flux, which yields the
desirable significance of 5$\sigma$ one can scale Crab flux as
\begin{equation}
F^{min}(>E_0) = 5(S/N)^{-1}\cdot F^{Crab}(>E_0)
\end{equation}
For the real detection of the source one has to control in
addition a number of detected $\gamma$-ray showers, which should
not be less than 10, which sets the second condition of the
minimum detectable flux
\begin{equation}
F^{min}(>E_0) \geq 10 (N_\gamma)^{-1} \cdot  F^{Crab}(>E_0)
\label{second}
\end{equation}
Note that for the detector considered here the first condition is
always stronger, which means that the total number of detected
$\gamma$-rays is in fact high, and the second condition is almost
always holds ($N_\gamma>>10$). The final results on the
sensitivity calculations for the system of two 30~m telescopes are
summarized in Table~\ref{flux}.

\begin{table}[t]
\caption{The integral rates (after cuts) and the minimum
detectable $\gamma$-ray flux (above $E_{th}$) after 50 hrs of
observations at 5$\sigma$ level with a system of two 30~m imaging
atmospheric \v{C}erenkov telescopes. \label{flux}}
\begin{center}
\begin{tabular}{lllll} \hline
$E_{th}$, GeV & $R_\gamma$, Hz & $R_{e}$, Hz & $R_{CR}$, Hz &
$F_{min}(>E_{th}),\, \rm cm^{-2}s^{-1}$ \\ \hline
5 & 5.5 & 2.5 & 1.0 & $6.45\times 10^{-11}$ \\
10 & 4.7 & 1.5 & 1.0 & $3.00\times 10^{-11}$ \\
30 & 2.5 & 0.18 & 0.90 & $1.05\times 10^{-11}$ \\
50 & 1.7 & 0.06 & 0.68 & $6.65\times 10^{-12}$ \\
100 & 1.0 & 0.01 & 0.34 & $3.19\times 10^{-12}$ \\ \hline
\end{tabular}
\end{center}
\end{table}

It is worth noting that at low energies the electron background is
dominant over the cosmic ray background. The electron component of
the primary cosmic rays at energies about 5 GeV becomes, in fact, a
major background. The images of electron and $\gamma$-ray showers
are almost indistinguishable in shape, which make the electron
background an absolute limiting factor for the detector
sensitivity. Improved angular resolutions might help to reject
contamination of the isotropic electron flux in observations of
the point-like $\gamma$-ray source.

The sensitivity of the system of two 30~m telescopes can be
compared to the corresponding sensitivities of the single
stand-alone telescope of the same size as well as with the array
of five such telescope operating in the stereoscopic mode - {\sc
STEREO ARRAY}. The signal and background rates, as well as
resulting sensitivities are given in
Table~\ref{flux1}\footnote{Note that for any of given
configurations the peak in the differential detection rate lays
well above the chosen energy threshold -- $E_o$=5~GeV.}.

\begin{table}[!t]
\caption{The integral $\gamma$-ray rates and background rate
(before applying the analysis cuts), the sensitivities for a
single stand-alone telescope (I), the system of two (II) and five
identical telescopes (V) above the energy threshold of 5 GeV.
\label{flux1}}
\begin{center}
\begin{tabular}{llll} \hline
Detector: & $R_\gamma$, Hz & $R_{b}$, Hz & $F_{min}(>E_{o}),\, \rm
cm^{-2}s^{-1}$
\\ \hline
I & 16.8 & 1726 & $1.46\times 10^{-10}$ \\
II & 11.2 & 1098 & $6.45\times 10^{-11}$ \\
V &  20.1 & 3261 & $2.94\times 10^{-11}$ \\ \hline
\end{tabular}
\end{center}
\end{table}

One can see in Table~\ref{flux1} that single ´stand-alone´
telescopes provide rather high $\gamma$-ray rates. However, due to
limited rejection power the final sensitivity is by more than a
factor of 2 lower than for a system of two telescopes, and by a
factor of 5 as compared with the system of five telescopes. At the
same time the increase in sensitivity with two to five telescopes
is relatively modest. It is provided by the increase in collection
area of the $\gamma$-ray showers, as well as a slightly better
angular resolution for higher telescope multiplicity events.

Given the results obtained here one can consider an alternative
approach, which is a construction of a few systems of two 30~m
telescopes spatially separated in the observational plane. In such
a case the sensitivity of entire telescope array will be
proportional to $n^{-1/2}$, where $n$ is a number of independent
two telescopes subsystems.

\subsection{Array Flexibility}

The flexibility in operation of the array of the atmospheric
imaging \v{C}erenkov telescopes is in fact an important issue. For
instance one can use two subsystems of the array for simultaneous
observations of two different sources. Apparently it increases the
observational time by factor of two, but with a somewhat reduced
sensitivity. On other hand in some cases all telescopes of the
array can be used independently in a stand-alone mode for the
monitoring of a number of potential $\gamma$-ray sources (like
AGNs) to perform quick-look discovery observations for the sample
of promising candidates.

\section{Conclusion}

Based on the detailed Monte Carlo simulations we have studied the
performance of the system of two, as well as five, 30~m imaging
atmospheric \v{C}erenkov telescopes. An array of five 30~m
telescopes is very close to the optimum detector for the
$\gamma$-ray observations above 10~GeV. Such an array of detectors allows
for the reduction of the energy threshold down to a few GeV, and would be the
most sensitive instrument in the field of ground-based high energy
$\gamma$-ray astronomy in this energy range.

The {\sc STEREO ARRAY} might be considered as a prototype for a
major future detector for ground-based high energy $\gamma$-ray
astronomy. Its construction and operation will enable scientists
to perform high quality $\gamma$-ray observations from the ground,
which will provide an overwhelming insight into the understanding
of the mechanics of high energy $\gamma$-ray emission from the
various physical environments in the Universe. The physics
rationale of the {\sc STEREO ARRAY}, provided with a sensitivity
value as given here, is a subject for another forthcoming paper.

\section*{Acknowledgements}

I would like to thank James Buckley, Werner Hofmann, and Trevor
Weekes for the discussions on a subject of this paper. I would
like also thank the referee, who remains anonymous, for comments
and suggestions, which have improved the quality of the paper.

\section*{References}

\begin{enumerate}
{\small \item Aharonian~et~al. {\bf The energy spectrum of TeV
gamma rays from the Crab Nebula as measured by the HEGRA system of
imaging air \v{C}erenkov telescopes}, {\it  ApJ}, 539, 317, (2000)
\item Aharonian~et~al. {\bf The Crab Nebula and Pulsar between 500
GeV and 80 TeV: Observations with the HEGRA Stereoscopic Air
Cerenkov Telescopes}, {\it ApJ}, Volume 614, Issue 2, pp. 897
(2004) \item Aharonian, F., Hofmann, W., Konopelko, A., V\"{o}lk,
H.J. {\bf The potential of the ground based arrays of imaging
atmospheric Cherenkov telescopes}, {\it Astroparticle Physics},
Volume 6, Issue 3-4, 343; 369; (1997) \item Aharonian, F.,
Konopelko, A., V\"{o}lk, H.J., Quintana, H.{\bf 5@5 - a 5 GeV
energy threshold array of imaging atmospheric Cherenkov telescopes
at 5 km altitude}, {\it Astroparticle Physics}, Vol. 15, N4, 335
(2001) \item Akhperjanian, A., Sahakian, V. {\bf Performance of a
20~m diameter v{C}erenkov imaging telescope}, Astroparticle
Physics, Volume 21, Issue 2, p. 149 (2004) \item Atoyan, A.,
Aharonian, A. {\bf On the mechanisms of gamma radiation in the
Crab Nebula}, {\it MNRAS}, 278, 525 (1996) \item Bernloehr et al.
{\bf The optical system of the H.E.S.S. imaging atmospheric
Cherenkov telescopes}, {\it Astroparticle Physics}, Vol. 20, Issue
2, 111 (2003) \item de~Jager et al. {\bf Gamma-Ray Observations of
the Crab Nebula: A Study of the Synchro-Compton Spectrum}, {\it
ApJ}, 457, 253 (1996) \item DuVernois~et~al. {\bf Cosmic-ray
electrons and positrons from 1 to 100~GeV: measurements with HEAT
and their interpretation}, {\it ApJ}, 559, 296 (2001) \item
Hinton, J. {\bf The status of the HESS project}, {\it New
Astronomy Reviews}, Vol. 48, Issue 5-6, 331 (2004) \item Hofmann,
W. {\bf How to focus a Cherenkov telescope}, {\it J. Phys. G:
Nucl. Part. Phys.} 27, No 4, 933 (2001) \item Konopelko, A., et
al. {\bf Performance of the stereoscopic system of the HEGRA
imaging air Cerenkov telescopes: Monte Carlo simulations and
observations}, {\it Astroparticle Physics}, Volume 10, Issue 4, p.
275 (1999) \item Konopelko, A. {\bf Space-angular and temporal
parameters of Cherenkov light emission in air showers of energy
from 1 TeV down to 10 GeV}, {\it Proc. of the Kruger Park Workshop
on TeV Gamma-Ray Astrophysics "Towards a Major Atmospherics
Cherenkov Detector - V}, (ed. O.C. de Jager), Kruger Park, South
Africa, August 8-11, 208(1997)

\item Konopelko, A. {\bf Altitude effect in Cerenkov light flashes
of low energy gamma-ray-induced atmospheric showers}, {\it Journal
of Physics G: Nuclear and Particle Physics}, Volume 30, Issue 12,
pp. 1835 (2004). \item Konopelko, A., Chlingarian, A., Reimer, A.,
{\bf Study on background rejection and energy resolution with a 30
m stand-alone imaging atmospheric Cherenkov telescope using
non-parameteric multi-variate methods in a sub-100 GeV energy
range}, in preparation (2005)
 \item Lorenz, E.,
Mirzoyan, R. {\bf What have we learned from the MAGIC telescope
developments for a future large IACT ?}, {\it Proc. Intern. Symp.
``High Energy Gamma-ray Astronomy'', Eds. F.Aharonian and H.J.
V\"{o}lk, AIP Conf. Proceed. 558}, 586 (2000) \item Lucarelli, F.,
Konopelko, A., Aharonian, F., Hofmann, W., Kohnle, A., Lampeitl,
H., Fonseca, V. {\bf Observations of the Crab Nebula with the
HEGRA system of IACTs in convergent mode using a topological
trigger}, {\it Astroparticle Physics}, Vol. 19, Issue 3, 339
(2003) \item MAGIC Proposal: {\bf ``The MAGIC Telescope'': design
study for the construction of a 17~m \v{C}erenkov telescope for
gamma-ray astronomy above 10~GeV} {\it MPI-PhE/98-5} (1998)\item
Masterson, C. {\bf Observation of galactic TeV gamma ray sources
with H.E.S.S.}, Proc. 28th Int. Cosmic Ray Conf., Tsukuba, Univ.
Academy Press, Tokyo, p. 2323 (2003)
 \item Sanuki et al. {\bf Precise measurement of
cosmic-ray proton and Helium spectra with the BESS spectrometer},
{\it ApJ}, 545, 1135 (2000) \item Simpson, J.A. {\it Ann. Rev.
Nucl. and Part. Sci.}, 33, 323 (1983) \item Wiebel, B. {\bf
Chemical composition in high energy cosmic rays}, {\it Fachbereich
Physik Bergische Iniversit\"{a}t Gesamthochschule, Wuppertal, WUB
94-08} (1994) }
\end{enumerate}

\end{document}